# Comprehensive Efficient Implementations of ECC on C54xx Family of Low-cost Digital Signal Processors

Muhammad Yasir Malik

`yasir_alf@yahoo.com`

**Abstract.** Resource constraints in smart devices demand an efficient cryptosystem that allows for low power and memory consumption. This has led to popularity of comparatively efficient Elliptic curve cryptography (ECC). Prior to this paper, much of ECC is implemented on reconfigurable hardware i.e. FPGAs, which are costly and unfavorable as low-cost solutions.

We present comprehensive yet efficient implementations of ECC on fixed-point TMS54xx series of digital signal processors (DSP). 160-bit prime field $GF(p)$ ECC is implemented over a wide range of coordinate choices. This paper also implements windowed recoding technique to provide better execution times. Stalls in the programming are minimized by utilization of loop unrolling and by avoiding data dependence. Complete scalar multiplication is achieved within *50 msec* in coordinate implementations, which is further reduced till *25 msec* for windowed-recoding method. These are the best known results for fixed-point low power digital signal processor to date.

## 1 Introduction

No other information security entity has been more extensively studied, researched and applied as asymmetric cryptography, thanks to its ability to be cryptographically strong over long spans of time. Asymmetric or public key cryptosystems (PKC) have their foundations in hard mathematical problems, which ensure their provable security at the expanse of implementation costs. So-called "large numbers" provide the basis of asymmetric cryptography, which makes their implementation expensive in terms of computation and memory requirements.

Numerical and computative dilemmas like factoring, discrete logarithm, decoding of linear codes and quantum mechanical effect problems are all being used in public key cryptography. These phenomenons have led the construction of many public-key algorithms; RSA, Diffie-Hellman, El-Gamal encryp-tion scheme, elliptic curve cryptography and McEliece cryptosystem are most commonly names amongst many others.

Public key cryptography enjoys utilization in many security applications thanks to its ability to render various security assurances. In present day, they are an essential security segment of almost all the operations involving two or more parties. Their wide acceptance leads to creation of efficient and secure protocols that accommodate near-to-perfect protection. On a lower note, in spite of huge developments in the capacity and processing capabilities of embedded systems, asymmetric cryptography is still considered a scheme for "big boys". This is a point of concern for researchers that many interesting applications i.e. smart cards, car security and wireless sensor networks are not fully able to reap the benefits of public key deployment. However hybrid infrastructure, which is combination of symmetric and asymmetric parameters, is in place in many appliances.

## 1.1   Comparison of RSA and ECC

Of all the above mentioned public key cryptosystems, elliptic curve cryptography (ECC) is of utmost interest due to fewer bits involved in its operation, making it an attractive choice for a variety of applications. Use of PKC on smart devices with restricted resources can be materialized by utilizing ECC. Small key sizes of ECC provide the same level of security as conventional cryptosystems which are based on integer factorization and discrete logarithmic problems. Ratio of security delivered to the key size of ECC is also larger than RSA or other cryptosystems, indicating better security of ECC for given key size. A comparison of equally secure cryptosystems is given in Table 1.

**Table 1.** Comparison of RSA and ECC

| RSA key length | ECC key length | Ratio | MIPS year |
|---|---|---|---|
| 512 | 106 | 5:1 | $10^4$ |
| 768 | 132 | 6:1 | $10^8$ |
| 1024 | 160 | 7:1 | $10^{11}$ |
| 2048 | 210 | 10:1 | $10^{20}$ |

## 1.2   ECC Variants

Although elliptic curves have been in literature for over 150 years, there use for cryptography was suggested by Miller and Koblitz around twenty-five years ago. ECC employs elliptic curve discrete logarithm problem (ECDLP), which is similar to the discrete logarithm problem (DLP). By exploiting this similarity, nearly all the DLP-applying cryptosystems over integer modulo $p$ can also be accomplished in elliptic curves. This renders new algorithms like ECDH, ECDSA and ECMQV etc based on the principles of Diffie-Hellman key exchange protocol, digital signature scheme DSA and authenticated protocol of key agreement, MQV.

## 1.3 Applications of ECC

The described factors call for wider role of ECC in present-day cryptography, specifically in resource-starved applications thanks to its faster implementations and low power and bandwidth consumptions. Trust in ECC security cause inclusion of many ECC protocols in internationally distinguished protocols such as IEEE Std 1363-2000, ISO CD 14888-3, ANSI X9.62 etc.

In last ten years, there has been tremendous work on the different ways and methodologies to enhance the implementation capabilities of ECC. Many novel approaches are introduced to speed up either overall system or the critical operation of scalar multiplication to find *kP* for any given point *P* on the curve $F_p$. Scalar multiplication is most significant process in ECC, and can account for about 70-80% of the total computations. In spite of the volume of research into ECC, it is interesting to note however, that the work done on the hardware and especially on the less expensive platforms is still sparse.

In the light of the fact that many appliances now need PKC to impose security in their structure, there is an ever-increasing demand of less expensive lightweight crypto hardware. Vehicle security, wireless sensor networks and many more exciting fields like these need asymmetric cryptography in their ranks. Due to resource limitation, these applications have to apply either hybrid cryptography (combination of symmetric and asymmetric cryptography) or only symmetric cryptosystems. Even though special-purpose PKC hardware can be utilized for arranging asymmetric cryptosystems in many applications, their implementation on less costly hardware is still a dilemma.

## 1.4 ECC on Hardware

While implementations of ECC are available on 8-bit controllers and ASICs and FPGAs, many of these attempts don't claim to be efficient in terms of computation. Our work presents implementation of ECC on 16-bit low cost, low power digital signal processor (DSP). DSPs are the most suitable candidates for secure hardware deployments, and they are already serving this purpose in many fields of application. DSPs have more computational capabilities and storage space than basic 8-bit μcontrollers, while being low power in nature. Another advantage of using DSPs as crypto-hardware is their relatively low cost as compared to FPGAs, which are believed to provide more effi-cient ECC operations but lack major support in deployment because of high expense. Till date, few works have been done on ECC for DSPs, with our contribution being at the forefront.

## 1.5   Our Contributions

In this work, we will address following points.
— Present different implementation choices of ECC for TMS320C54xx, a very popular family of DSPs being manufactured by Texas Instruments.
— Implement ECC for different coordinates including the newly introduced Edward and Inverse Edwards curves.
— Use of efficient programming techniques to optimize performance of ECC.
— Point out programming techniques for efficient performance of ECC.
— Implement a new windowed-NAF of scalar to enhance the capability and performance of hardware.
— Compare our results with previously performed research works on ECC.

## 2   Related Works

Comprehensive details of cryptosystem implementations on different categories of hardware can be found in [1] and [2]. Several cryptosystems have been restructured and exercised on DSPs in the past, with particular emphasis on RSA. Implementations concerning elliptic curve cryptography (ECC) and its components on different platforms are also in literature.

As mentioned earlier, last decade saw an expansion in research on ECC in terms of hardware actualization. Many new hardware architectures and calculation-saving techniques are developed, to make ECC realizable in practice. The platforms put to use for this objective vary from simple [3], in which authors use an 8-bit Atmel ATmega128 µcontroller, operating at 8MHz. They were able to achieve ECC over prime field $GF(2^{160})$ in less than a second. Software and hardware co-design for a µcontroller is presented in [4], using Dalton 8051 and special hardware accelerator composed of an elliptic curve acceleration unit (ECAU) and an interface with direct memory access (DMA) to enable fast data transfer between the ECAU and the external RAM (XRAM) attached to the 8051 µcontroller. ECAU accomplished full scalar multiplication over prime field $GF(2^{191})$ in 118 msec, when Dalton 8051 is clocked at 12MHz.

Relatively more implementation work on ECC has been performed using FPGAs, which allow reconfigurable design realizations. Some of such reconfigurable designs are presented in [5] and [6]. Elliptic curve designs over prime field $GF(p)$ are given in [7,8,9], in which architectures are based on systolic array Montgomery modular multipliers. Many other papers are also available for reconfigurable-hardware designs [10], LD algorithm [26] and the configuration by use of Karatsuba multipliers to name a few. Authors in [10] present efficient ECC over NIST primes P-224 and P-256 on commercial Xilinx'sVirtex-4SX55 FPGA.

In the line of DSP platforms, the earliest works are featured in [11, 12, 13, 14, 15]. Montgomery multiplication algorithm is a popular method to improve speed of modular multiplications in public-key cryptosystems. Work in [11] suggests enhanced Montgomery multiplication based on DSP architectures. An efficient processor over prime field $GF(p)$ implemented on DSPTMS320C6201 is proposed in [12]. It uses a Montgomery multiplier with a 16-bit digit size, which can be used for carrying out RSA, DSA and ECDSA, and elliptic doubling to increase speed. Authors in [13] present processor using MAC2424 ($24 \times 24$ -bit) for $GF(p)$, taking into account the precomputations required by the signed window scalar multiplication method. In [14], a windowing mechanism by combining NAF and variable-length sliding window, used in implementation of elliptic curve digital signature on TMS320VC5402, is given to decrease the complexity of point multiplication. ECC and its components over $GF(2^{160})$ are implemented on fixed point TMS320VC5416 DSP in [15], achieving scalar multiplication in 63.4 msec.

## 3  Preliminaries

In this section, we briefly describe ECC, the systems of coordinates used by it and DSP which we use for implementations.

### 3.1 Elliptic Curves

**Definition 1.** An elliptic curve E over a field $K$ is defined by an equation

$$E: y_2 + a_1 xy + a_3 y = x_3 + a_2 x_2 + a_4 x + a_6 \qquad (1)$$

where $a_1, a_2, a_3, a_4, a_6 \in K$, and $\Delta \neq 0$, where $\Delta$ is the discriminant of $E$ and must be non-zero to ensure no point on the curve has two or more distinct tangent lines.

This equation is known as Weierstrass equation, where $(x, y)$ tuple represents the point on the curve. Point $O$ at infinity also satisfies the projective form of Weierstrass equation and serves as identity for the abelian group formed by set of points $E(K)$ on the curve. Adding the points on the elliptic curve $E$ is the fundamental group operation and for two points $P(x_1, y_1)$ and $Q(x_2, y_2)$, their addition $R\ (x_3, y_3)$ can be given as $R\ =\ P\ +\ Q$. On the curve, tangent-and-chord process is applied to perform this operation, with two defined cases of $P\ =\ Q$ and $P\ \neq\ Q$. In the first case, point doubling is carried out while point addition is done in the later case. Scalar multiplication is the basic crypto function of ECC, and it uses point addition and point doubling to find $kP$, which is the result of adding the point $P$ to itself $k$ times

$$kP\ =\ P\ +\ P\ +\ \cdots\ +\ P\ (k \text{ times}).$$

The length of $k$ depicts the key size of corresponding ECC cryptosystem, and can be in the range of 160-521 bits.

Security of ECC relies on the elliptic curve discrete logarithm problem (ECDLP), defined as follows:

**Definition 2.** Given the prime modulus $p$, the curve constants $a$ and $b$ and two points $P$ and $Q$, find a scalar $k$ such that $Q\ =\ kP$, where $kP$ is the scalar multiplication as defined above.

The equation (1) can be reduced by admissible changes of variables. If the characteristic of $K$ is not equal to 2 and 3, then equation (1) is rewritten as

$$y_2\ =\ x_3\ +\ ax\ +\ b, \qquad (2)$$

where $a, b\ \in\ K$, and $\Delta =\ 4a_3\ + 27b_2\ \neq\ 0$. This equation represents Weierstrass equation in its simple form, and is the most commonly used representation of ECC.

Points on curve $E(GF(2^m))$ when shown according to two-dimensional entities $(x, y)$, such as $x, y\ \in\ GF(2^m)$ are said to be using affine coordinates. Although they are the most basic form of curves, they exercise expensive inversion operation over prime fields. Point addition and point doubling in these coordinates require one inversion process each, making the choice of affine coordinates too costly for practical utilization.

### 3.1.1   Jacobian Projective Coordinates

For efficient implementation of scalar multiplications, these points are converted into three-point coordinates such as Jacobian projective coordinates, with the curve projective equation $Y_2\ =\ X_3\ 3XZ_4\ +\ bZ_6$ and point at infinity as $(1,1,0)$. Each point in affine coordinate representation $(x, y)$ can be represented by a reciprocal form in projective Jacobian coordinates $(X, Y, Z)$ such that,

$$P(x; y)\ \equiv\ P(X; Y; Z)$$

The formulae for conversion between corresponding coordinates are:

**Affine-to- Jacobian projective**: $X = x; Y = y; Z = 1$
**Jacobian projective-to-Affine**: $x = X/Z_2; y = Y/Z_3$, implying $Z \neq 0$.

Various representations of operations on point addition and point doubling are in literature, however we chose most efficient for our implementations. This provides clear insight into the performance improvements taking into account the recent developments in the algorithms.

Let $P(X_1, Y_1, Z_1)$ and $Q(X_2, Y_2, Z_2)$ be two points in Jacobian coordinates such that $P \neq Q$ then $R = P + Q$, where $R(X_3, Y_3, Z_3)$ is the point addition of $P_1$ and $P_2$, is given as:

$$A \leftarrow Y_1^2, \quad B \leftarrow 4X_1 A, \quad C \leftarrow 8A^2, \quad D \leftarrow 3X_1^2 + aZ_1^4,$$
$$X_3 \leftarrow D^2 - 2B, \quad Y_3 \leftarrow D(B - X_3) - C, \quad Z_3 \leftarrow 2Y_1 Z_1$$

where $A, B, C$ and $D$ are subordinates for storing intermediate values.

Similarly, for $P = Q$ the point doubling $R = 2P$ is obtained as:

$$A \leftarrow Z_1^2, \quad B \leftarrow Z_1 A, \quad C \leftarrow X_2 A, \quad D \leftarrow Y_2 B, \quad E \leftarrow C - X_1,$$
$$F \leftarrow D - Y_1, \quad G \leftarrow E^2, \quad H \leftarrow GE, \quad I \leftarrow X_1 G,$$
$$X_3 \leftarrow F^2 - (H + 2I), \quad Y_3 \leftarrow F(I - X_3) - Y_1 H, \quad Z_3 \leftarrow Z_1 E$$

Both of these operations are in simple form, using more variables than required, and can be further reduced by eliminating the repeated calculations and through optimized choice of a, in the elliptic curve equation.

### 3.1.2 Lopez-Dahab(LD) Coordinates

Lopez-Dahab (LD) are another popular coordinate system with the projective elliptic curve equation,
$$Y_2 + XYZ = X_3 Z + aX_2 Z_2 + bZ_4$$
and point of infinity at (1,0,0).

(LD) coordinates are preferable instead of affine, and sometimes Jacobian coordinate due to lesser field multiplications to perform a point addition and their flexibility of computing the point addition in mixed coordinates i.e. adding points represented in distinct coordinate systems.
The formulae for point conversion between affine and LD projective coordinates are:

**Affine-to-LD projective**: $X = x; Y = y; Z = 1$
**LD projective -to-Affine**: $x = X/Z; y = Y/Z_2$, implying $Z \neq 0$.

Point addition $R\ (X_3,\ Y_3,\ Z_3)$ of two points $P\ (X_1,\ Y_1,\ Z_1)$ and $Q\ (X_2,\ Y_2,\ Z2$ can be shown as, as provided in [16]:
$$A_1 = X_1 Z_2, \quad A_2 = X_2 Z_1, \quad C = A_1 + A_2,$$
$$B_1 = A_1^2, \quad B_2 = A_2^2, \quad D = B_1 + B_2,$$
$$E_1 = Y_1 Z_2^2, \quad E_2 = Y_2 Z_2^1, \quad F = E_1 + E_2,$$
$$G = CF, \quad Z_3 = Z_1 Z_2 D,$$
$$X_3 = A_1(E_2 + B_2) + A_2(E_1 + B_1),$$
$$Y_3 = (A_1 G + E_1 D)D + (G + Z_3)X_3.$$
Simple point addition in this coordinate system takes $13M + 4S + 9A$, where $M$ denotes a multiplication, $S$ represents a squaring and $A$ is an addition.

Mixed addition requires $8M + 5S + 8A$ and one multiplication by $a$, as given in [17].
$$U = Z_2^2 Y_1 + Y_2, S = Z_2 X_1 + X_2, T = Z_2 S,$$
$$Z_3 = T^2, V = Z_3 X_1, C = X_1 + Y_1,$$
$$X_3 = U^2 + T(U + S^2 + a_2 T),$$
$$Y_3 = (V + X_3)(TU + Z_3) + Z_3^2 C.$$

Author in [18] gives doubling of point $P$, $2P = (X_3, Y_3, Z_3)$ that requires $4M + 4S + 5A$ and one multiplication by a, and is given by
$$S = X_1^2, \quad U = S + Y_1, \quad T = X_1 Z_1, \quad Z_3 = T^2, \quad T = UT,$$
$$X_3 = U^2 + T + aZ_3, \quad Y_3 = (Z_3 + T)X_3 + S^2 Z_3.$$

### 3.1.3 Edwards curves

In [19], author proposes new normal form of elliptic curves with an addition law symmetric in $x$ and $y$ coordinates. All elliptic curves can be transformed to this normal form by field extension.

An Edwards curve, over a field $k$ is given by the equation,
$x_2 + y_2 = 1 + dx_2 y_2$, where $d \in k \setminus \{0, 1\}$.

Edwards addition law for adding any two points $(x_1, y_1)$ and $(x_2, y_2)$ on this curve is shown as
$$(x_1, y_1), (x_2, y_2) = \left(\frac{x_1 y_2 + y_1 x_2}{1 + dx_1 x_2 y_1 y_2}, \frac{y_1 y_2 - x_1 x_2}{1 - dx_1 x_2 y_1 y_2}\right).$$

Neutral element of this addition is $(0, 1)$ and the inverse of any point $(x_1, y_1)$ on the curve $E$ is $(-x_1, y_1)$ On Edwards curves, doubling is performed by using the same formula as that of addition.

The curve equation can be set to $(X_2 + Y_2)Z_2 = C_2(Z_4 + dX_2Y_2)$, with the identity at $(0, c, 1)$.

Any point $(X_1, Y_1, Z_1)$ following the above curve equation with condition $Z_1 \neq 0$, coincides to the affine point $(X_1/Z_1, Y_1/Z_1)$.

Point addition $(X_3, Y_3, Z_3)$ of two points with coordinates at $(X_1, Y_1, Z_1)$ and $(X_2, Y_2, Z_2)$ is shown as:
$$A = Z_1 Z_2, \quad B = A_2, \quad C = X_1 X_2, \quad D = Y_1 Y_2, \quad E = dCD,$$
$$F = B - E, \quad G = B + E,$$
$$X_3 = AF\big((X_1 + Y_1)(X_2 + Y_2) - C - D\big),$$
$$Y_3 = A G (D - C), \quad Z_3 = cFG.$$
It takes $10M + 1S + 1C + 1D + 7a$ computations.

Doubling of a point $(X_3, Y_3, Z_3) = 2(X_1, Y_1, Z_1)$, uses $3M + 4S + 3C + 6a$.
$$B = (X_1 + Y_1)^2, \quad C = X_1^2, \quad D = Y_1^2, \quad E = C + D,$$
$$H = (cZ_1)^2, \quad J = E - 2H, \quad X_3 = c(B - E)J,$$
$$Y_3 = cE(C - D), \quad Z_3 = EJ.$$

### 3.1.4 Inverted Edwards Curves

Inverted Edward curves represent an affine point $(X_1 : Y_1 : Z_1)$ as $(Z_1/X_1, Z_1/Y_1)$. The curve equation to represent any point on the Edward curve is,
$$(X_{12} + Y_{12})Z_{12} = X_{12} Y_{12} + dZ_{14}$$
satisfying the condition $X_1 Y_1 Z_1 \neq 0$.

Computing $(X_1 Z_1 : X_1 Z_1 : X_1 Y_1)$ converts standard Edwards coordinates $(X_1 : Y_1 : Z_1)$ to inverted Edwards coordinates and vice versa.

Point addition $(X_3 : Y_3 : Z_3)$ of two points with coordinates $(X_1 : Y_1 : Z_1)$ and $(X_2 : Y_2 : Z_2)$ takes $9M + 1S + 1D + 7a$ computations, and is done as:
$$A = Z_1 Z_2, \quad B = dA^2,$$
$$C = X_1 X_2, \quad D = Y_1 Y_2, \quad E = C D,$$
$$H = C - D, \quad I = (X_1 + Y_1)(X_2 + Y_2) - C - D,$$
$$X_3 = (E + B)H, \quad Y_3 = (E - B)I, \quad Z_3 = A H I.$$

Procedure for point doubling in inverted Edwards curve is:
$$A = X_1^2, \quad B = Y_1^2, \quad C = A + B, \quad D = A - B,$$
$$E = (X_1 + Y_1)2 - C,$$
$$Z_3 = D.E, \quad X_3 = C.D, \quad Y_3 = E(C - 2dZ_1^2),$$

and takes $3M + 4S + 1D + 6a$.

If doubling is followed by addition in inverted Edwards curve, deploying point tripling offers reduced calculations and resources. The following calculations take $7M + 7S + 1D + 17a$.

$$A = X_1^2, \quad B = Y_1^2, \quad C = Z_1^2, \quad D = A + B,$$
$$E = 4(D - dC),$$
$$H = 2D(B - A), \quad P = D^2 - AE, \quad Q = D^2 - BE;$$
$$X_3 = (H + Q)((Q + X_1)^2 - Q^2 - A), \quad Y_3 = 2(H - P)P\, Y_1,$$
$$Z_3 = P((Q + Z_1)^2 - Q^2 - C).$$

### 3.2 Scalar representation

Most efficient scalar multiplication algorithms follow Horner polynomial representation:

$$a_n x^n + a_{n-1} x^{n-1} + \cdots + a_2 x^2 + a_1 x^1 + a_0$$
$$= a_0 + (a_1 + (a_2 + \ldots + a_{n-1} + a_n + x)x)\ldots)x)x)x$$

**Definition:** Let the scalar $k$ is in the form,

$$\sum b_j 2^j = b_n 2^n + b_{n-1} + 2^{n-1} \ldots$$

then the scalar multiplication $kP$ can be calculated as,

$$kP = \sum_{j=0}^{m-1} b_j 2^j P = 2(\ldots 2(2b_m - 1P + b_m - 2P) + \ldots) + b_0 P.$$

Hamming weight (the number of one's) in such representation is nearly half of total number of bits, $m/2$ where m is the length of the scalar $k$ and can be from 160 *to* 571 bits for practical systems. Running time of this procedure is approximately,

$$m/2A + mD.$$

$A$ and $D$ represent number of point additions and doubling respectively.

#### 3.2.1 Windowed Non-Adjacent Form (*w*NAF)

**Definition**: A non-adjacent form of a scalar $k$ is given by expression $k = \sum_{i=0}^{l-1} u_i 2^i$ where non-zero constant $u_i$ is odd, and there are no adjacent non-zero digits, for length $m$.

**Definition:** For width $w > 1$, we can expand any positive integer $k$ using

NAF of width $w$, such as
$$wNAF(k) = \{u_{l-1}, \ldots, u_0\}$$

The running time for $w$-NAF is
$$\frac{m}{(w+1)}A + mD.$$

### 3.2.2  New method for windowed-recoding

In [21], a new technique for recoding of scalar $k$ based on one's complement is given. In this work, we will elaborate the method and implement it, something which has not been done before.

**Definition:** We can write one's complement of any number $N$ with bit length $m$ as
$$C1 = (2m - 1) - N.$$
Reordering of this equation gives
$$N = 2m - C1 - 1.$$

This representation can be windowed for any width $w$, $w > 1$ and it provides the running time of
$$\frac{m}{(w+1)}A + (m-w)D.$$

One point addition costs $1I + 3M$, whereas the cost of a point doubling is $1I + 4M$. By using the proposed process, considerable computations can be saved as they offer more flexibility in terms of storage needs due to pre-computations and optimized performance based on the platform competence.

Table 2. Performance comparison of different point representations

| Point Representation | Length | #P A | #P D | Precomputation |
|---|---|---|---|---|
| binary | $m$ | $m/2$ | $m$ | --- |
| recoded binary | $m$ | $m/3$ | $m + 1$ | --- |
| w-NAF | $m$ | $m/(w+1)$ | $m + 1$ | Table of $(2^{w-1} - 1)$ m-bit multiples |
| w-NAF* | $m$ | $m/(w+1)$ | $m - w$ | Table of $(2^{w-1} - 1)$ m-bit multiples. |

## 4    54xx Family Digital Signal Processor

We bring into play fixed- point digital signal processor TMS320VC5416, which belongs to C54xx family of DSPs by Texas Instruments, for our implementations, analysis and comparisons. Its multi-bus architecture works with three 16-bit data memory buses and a program memory bus. It houses 40-bit Arithmetic Logic Unit (ALU) including a 40-bit barrel shifter and two independent 40-bit accumulators. Eight auxiliary registers **AR0~AR7** can be exploited during programming. The DSP accommodates 128k x 16-bit on-chip RAM and a $16k \times$ 16-bit on-chip ROM for program memory [27]. It offers 6.25-ns and 8.33-ns single-cycle execution times, which corresponds to their operations at 160 and 120 MIPS respectively.
Now we will describe some techniques to enhance performance of DSP.

### 4.1    Data Dependence

As we are using assembly for our implementation, the problems due to data dependence such as write after read (WAR), read after write (RAW) and write after write (WAW) may arise. Some variables need to be used multiple times and when it happens in consecutive instructions, data dependence may lead to errors in the program.

As shown in the following example, the value in register **AR2** is being used multiple times in consecutive instructions; firstly its value is loaded to accumulator **A**, then it is being multiplied by the value in **A** (squaring process). The result is again stored in the same register. This causes stalls because of the consumption of only one register (or variable) in chronological order.

```
DLD *AR2,A
MPYU *AR2,A
STL A,*AR2-
```

To remove the problems arising from data dependence, stalls (NOPs in programming) are inserted between the instructions. Scheduling remains the favored procedure to avoid these problems, in which order of execution of instructions is altered and thus less stalls are required.

### 4.2    Loop unrolling

To decrease the control dependence, loop is replicated several times. This process may increase the code length and memory storage requirements, but nonetheless helps in attaining better performance as the number of NOPs decreases.

```
for (i=0; i<100; i=i+1)
```

```
x[i] = x[i] + a;
```

The above statements can be loop unrolled as shown.

```
for (i=0; i<100; i=i+4) {
x[i] = x[i] + a;
x[i+1] = x[i+1] + a;
x[i+2] = x[i+2] + a;
x[i+3] = x[i+3] + a;
}
```

## 5  Simulations and Results

As discussed earlier, we use 160-bit over $Fp$ for implementation and examination. This section contains the simulation results for our implementations on C54x DSP, which is running at 160 MHz.

Table 3. show the improvement in the results for Jacobian coordinates by our procedure, as compared to the earlier work on the similar platform. Development achieved in our work here is mainly due the use of data dependence and loop unrolling. To keep the code length from being too lengthy and ambiguous, loops are unwound for five times at the maximum.
Restricted number of registers in C54x allows us to make use of memory to store variables, thus reducing reliance on registers and the data stored in them.

**Table 3.** Performance comparison of 160-bit elliptic curves over $Fp$ using Jacobian coordinates

|  | Previous work on 160-bit $Fp$, Jacobian coordinates [15] | | Our work on 160-bit $Fp$, Jacobian coordinates | | Improvement |
|---|---|---|---|---|---|
|  | CPU cycles | Time | CPU cycles | Time |  |
| Addition in $Fp$ | 315 | 1.97us | 274 | 1.71 $\mu s$ | 13.01% |
| Subtraction in $Fp$ | 357 | 2.23us | 320 | 2.0 $\mu s$ | 10.36% |
| Inversion | -- | -- | 16452 | 103 $\mu s$ | -- |
| Montgomery Multiplication | 2,860 | 17.88us | 2654 | 16.59 $\mu s$ | 7.2 % |
| Point Addition | 33,049 | 207 us | 28,769 | 179.8 $\mu s$ | 12.95% |
| Point Doubling | 40,737 | 254 us | 34,272 | 214.2 $\mu s$ | 15.87% |
| Scalar Multiplication | 10,148,863 | 63.4 ms | 8,796,721 | 54.98 ms | 13.28% |

Our work achieves considerable improvements in the previous work on the same curves and coordinates. Overall performance improves by an impressive *13.28 percent*, and time executed by complete scalar multiplication *kP* decreases to about *55 msecs*. It shows that efficient programming can be vital to overall contribution to an efficient cryptosystem.

Now we present the efficient implementations of ECC using other coordinate systems, including and using the recently developed results on Edwards and Inverted Edwards curves.

**Table 4.** 160-bit elliptic curves over $Fp$ over Lopez Dahab (LP)

|  | CPU cycles | Execution time | %age Improvement |
|---|---|---|---|
| Point addition | 37,604 | 235 $\mu$s | -13.8 % |
| Point doubling | 25,114 | 157 $\mu$s | 38.35 % |
| Scalar multiplication | 7,001,756 | 43.77 ms | 31.0 % |

**Table 5.** 160-bit elliptic curves over $Fp$ over Edwards curves

|  | CPU cycles | Execution time | %age Improvement |
|---|---|---|---|
| Point addition | 33448 | 209 $\mu$s | -1.2 % |
| Point doubling | 24,591 | 153.7 $\mu$s | 39.63 % |
| Scalar multiplication | 6,750,609 | 42.19 ms | 33.48 % |

**Table 6.** 160-bit elliptic curves over $Fp$ over Inverse Edwards curves

|  | CPU cycles | Execution time | %age Improvement |
|---|---|---|---|
| Point addition | 31,591 | 197.44 $\mu$s | 4.4 % |
| Point doubling | 22,506 | 140.66 $\mu$s | 44.75 % |
| Scalar multiplication | 6,105,734 | 38.16 ms | 39.84 % |

As obvious from the tables, the new curves present a favorable alternate to traditional Jacobian coordinates, and newer discoveries in finding best point addition and doubling methods are making them more efficient in terms of computations. Scalar multiplication takes only *38 msec* for Inverse Edward curves, which clearly shows their effectiveness in implementing ECC.
Efficient recoding algorithms can further enhance the performance at the cost of precomputations and memory storage for caching the multiples of elliptic point, each of them is 160 bit in length here.

Average Hamming weight of a w-NAF expression is $(w + 1) - 1$. Recoding algorithms decrease the point additions for scalar multiplication, whereas the number of point doubling may remain same. New recoding algorithm (Section 3.2.2) also reduces the point doublings in terms of $(m - w)$, for $w > 1$.
Table 2. show the parameters for the scalar multiplication performed on Inverse Edwards points to maximize the realization.

From Table 7. it is clear increasing the parameter improves the performance in terms of execution times. Times needed for scalar multiplication is reduced to *30.06*, *27.57* and *24.47 msecs* depending on the selection of window size.
But this improvement comes at the cost of pre-computations and memory. In this case, optimized parameters can be selected as per the need of applications.

**Table 7.** 160-bit elliptic curves over $Fp$ over w-NAF form

| Window size | # PA | #PD | Pre-computations | Memory space | CPU cycles | Execution time |
|---|---|---|---|---|---|---|
| w=3 | 40 | 157 | 7 | 3 x 16-bits | 4,810,438 | 30.06 ms |
| w=5 | 27 | 155 | 31 | 15 x 16-bits | 4,411,481 | 27.57 ms |
| w=10 | 15 | 150 | 1023 | 511 x 16-bits | 3,915,456 | 24.47 ms |

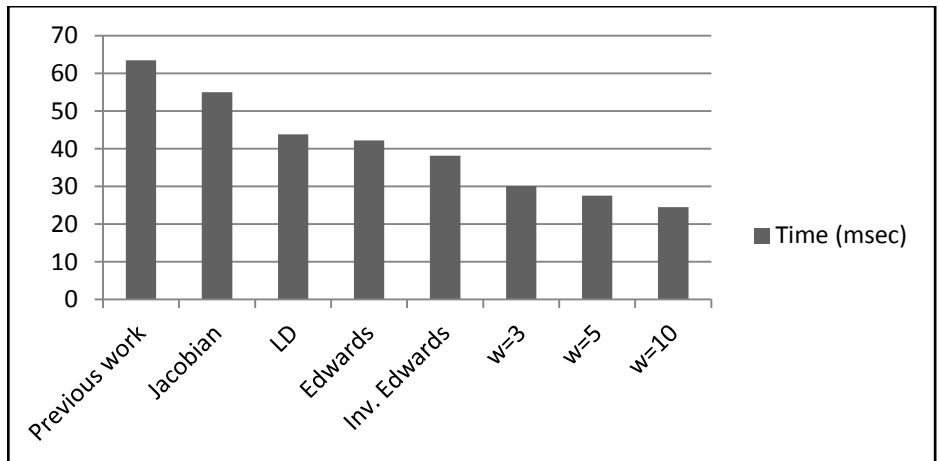

**Figure 1.** Scalar mulplications of 160-bit $Fp$ elliptic curves over different coordinates

In above figure, all implementations on DSP over different coordinates of size 160-bit ($F_p$) are shown.

## 6   Discussions

In this paper, we presented efficient implementations of ECC over $GF(2^{160})$, using all popular coordinate systems on a low cost and low power platform, on which not much work has been done in the past. We also showed the implementation results of a new windowed-recoding method, giving better results than any other general purpose fixed point DSP has done before. This comes at some storage cost, though DSP has enough memory to support pre-computations. Our work used efficient programming techniques of loop unrolling and anti-data dependence to ensure smooth operation. We make use of the low cost DSP to present efficient implementation with execution time within the range from *25 msec* to nearly *50 msec* for scalar multiplication over various coordinates. This work highlights general purpose DSPs as prospective element in future low cost secure systems and networks.